\documentclass[12pt]{article}

\usepackage{amsmath,amsthm,amssymb,color}
\usepackage{amsfonts,latexsym, hyperref, bbm, mathrsfs}
\newcommand{\brr}{({\bf r}-{\bf r}')}

\title{Noncommutative Geometry\\ and\\ Fluid Dynamics}

\begin{document}

{\bf \maketitle}

\vspace{0.0cm}
\begin{center}
	Praloy Das \footnote{E-mail: praloydasdurgapur@gmail.com}and
	Subir Ghosh \footnote{E-mail: subirghosh20@gmail.com }\\
	\vspace{2.0 mm}
	\small{\emph{Physics and Applied Mathematics Unit, Indian Statistical
			Institute\\
			203 B. T. Road, Kolkata 700108, India}} \\
\end{center}
\vspace{0.5cm}

\begin{abstract}
In the present paper we have developed a Non-Commutative (NC) generalization of perfect fluid model from first principles, in a Hamiltonian framework. The noncommutativity is introduced at the Lagrangian (particle) coordinate space brackets and the induced NC fluid bracket algebra for the Eulerian (fluid) field variables is derived. Together with a Hamiltonian this NC algebra generates the generalized fluid dynamics that satisfies exact local conservation laws for  mass and energy thereby maintaining mass and energy conservation. However, nontrivial NC correction terms appear in charge and energy fluxes. Other non-relativistic spacetime symmetries of the NC fluid are also discussed in detail. This constitutes the study of kinematics and dynamics of  NC fluid.

In the second part we construct an extension of Friedmann-Robertson-Walker (FRW) cosmological model based on the NC fluid dynamics presented here. We outline the way in which NC effects generate cosmological perturbations bringing in anisotropy and inhomogeneity in the model. We also derive a NC extended Friedmann equation.
\end{abstract}	

 \section{Introduction}
 It is very intriguing to learn that there are two parallel languages, fluid mechanics and Non-Commutative (NC) gauge theories, in which several completely unrelated physical systems can be formulated \cite{n0}. For example the discrete nature of Quantum Hall liquid can be described by NC gauge theories  \cite{n0}. On the other hand, the analogy between diffeomorphism and related transformations in the fluid  space and the Seiberg-Witten map \cite{sw} in NC gauge theories has led to a series of works by Jackiw and co-workers \cite{n01}, where these connections are studied from a deeper perspective. (For a lucid review see \cite{jac}). For example magnetohydrodynamics in very strong magnetic field, where excitations in the lowest Landau level play a dominant role, provides a natural framework of NC gauge theory \cite{gural,n01}. The connection between fluids and NC field theories was envisaged much earlier in \cite{or}.
 
 The above disconnected ideas have naturally led to attempts to formulate a generalization of canonical perfect fluid theory in NC space(time) where the space(time) is endowed with a noncommutative structure. Incidentally we have put the 'time' in parenthesis simply to indicate that conventionally the NC extension is restricted to spatial sector only without affecting the time.  A few relevant works in this context are \cite{van, mar}.

  In this perspective we have provided an explicit model of {\it{a generalized perfect fluid living in noncommutative space}}. We stress  that the model is  constructed from first principles, completely based on the map between the Lagrangian and Hamiltonian (Euler) formulation of fluid dynamics. The NC effect is induced in the Euler continuum algebra from the NC extension in discrete Lagrangian variable algebra. 
  
  There are two parallel ways of constructing NC  extension of a theory. One route, (if one is interested in NC field theory), is to directly apply the Groenwald-Moyal $*$-product \cite{sw,ncrev} in a conventional model field theory to obtain the NC-extended  field theory action. (See eg. earlier reviews \cite{ncrev} for generic NC field theory construction and $*$-product formalism in \cite{van} for NC fluid model. The other alternative, that is especially suited for fluids, is to start from the Lagrangian (discrete) fluid model, introduce the NC space effect and subsequently exploit the map that connects the Lagrangian framework to Euler Hamiltonian (continuum) framework to induce NC effects in the fluid field theory. We have followed the latter approach, the  primary reason being that NC generalization can be unambiguously done in the discrete variable setup (see for example \cite{mar}) {\footnote{Infact the first NC model proposed by Snyder \cite{sny} was in discrete framework and later on numerous NC extensions in quantum mechanical models have been proposed \cite{bertol}.}}.  In the present work we have provided the full NC extended algebra among the Euler degrees of freedom, ie. density and velocity field variables  and furthermore have studied briefly some consequences of it in cosmology{\footnote{It should be noted that in the review \cite{jac} an NC generalized fluid variable algebra has been outlined starting from the discrete Lagrangian setup  but it is not complete since only the NC extension of  density-density local algebra has been computed.}}. 
 
  Let us briefly discuss the recent importance of Hydrodynamics and noncommutative field theory in theoretical high energy physics. Hydrodynamics can be described in very general terms as a universal description of long wavelength physics that deals with low energy effective degrees of freedom of a field theory, classical or quantum. Interestingly it applies equally well at macroscopic and microscopic scales, examples being liquid drop model in early nuclear physics and quark-gluon-plasma produced at
  RHIC/LHC (in the former) and generic fluid models in cosmology (in the latter). The physical reasoning behind the success of fluid models is the reasonable assumption that at sufficiently high energy densities local equilibrium prevails in an interacting field theory so that local inhomogeneities are smoothed out enough to reduce to continuous fluid variables. Indeed the constitutive relations provide the bridge between the fundamental and fluid degrees of freedom. In the fluid picture one simply deals with the age old continuity equation and the Euler force equation, (in the most simple non-relativistic framework for an ideal fluid). Two of the most exciting and topical  areas of recent interest are cosmology and AdS-CFT correspondence \cite{ads} as applied to gauge-gravity duality \cite{gg} and both of them rely heavily on conventional fluid dynamics. Hence it will be very interesting to study the effects of non-trivial changes  in fluid equations, (at a fundamental level), in these areas. In the present paper we aim to outline how fluid dynamics in Non-Commutative (NC) space can affect cosmology by introducing cosmological perturbations. We postpone discussions on NC effects induced by  fluid dynamics on gauge-gravity duality to a future publication.

  A recently rediscovered effect that has impacted the theoretical physics community in a major way is the idea of introducing a NC spacetime and subsequently a NC generalization of quantum field theory. Seiberg and Witten \cite{sw}, in their seminal work, resurrected the (not so successful) NC spacetime introduced by Snyder \cite{sny}, when the former demonstrated that in certain low energy limits open string dynamics (with end points on $D$-branes) can be simulated by NC field theories where the NC parameter $\theta_{\mu\nu}$ is to identified with the anti-symmetric two-form field $B_{\mu\nu}$. This idea has led to a huge amount of literature and we refer to a few review works in \cite{ncrev}. In the present paper we develop a generalized fluid dynamics, compatible with NC space (since we work in non-relativistic regime), from first principles. One generally tends to avoid introducing NC between time and spatial degrees of freedom due to possible complications regarding unitarity in the quantum theory.
  
  Before getting down to the actual computation let us pause for a moment to comment on the justification of our scheme of introducing noncommutativity. Conventionally in NC quantum mechanics one simply postulates a generalization of the canonical phase space commutator algebra such that it reduces to the canonical commutators for the vanishing of NC parameter. From purely algebraic standpoint the only restriction is that the NC algebra has to satisfy the Jacobi identity (assuming that the canonical commutator algebra does). Subsequent restrictions will come from  physical implications of the extra NC terms. Clearly if the NC contribution in the commutator are (non-operatorial) c-number in nature the above condition is satisfied since the NC terms in the double commutators  trivially vanish (see for example \cite{bertol}). On the contrary for NC extensions of operatorial form (such as Snyder, $\kappa$-Minkowski, Generalized Uncertainty Principle induced algebra) the NC extensions are severely restricted in order to satisfy Jacobi identity (see for example \cite{gup}). 
  
  Let us now come back to the classical scenario that is of present interest. Just as classical Poisson algebra is elevated to quantum commutators via the correspondence principle in exactly identical fashion one can think of a classical counterpart of the NC quantum algebra.  Once again Jacobi identity, in the sense of double brackets, plays an essential role since, for a Hamiltonian system the  symplectic structure (brackets) has to obey the Jacobi identity.  However there is an added twist in the classical setup where sometimes it is possible to judiciously introduce constraints in a particular model under study such that the NC generalized algebra can be identified with the Dirac bracket algebra (see for example \cite{ncop}), in Dirac's Hamiltonian formulation of constrained system \cite{dirac}. The advantage is  that (i) one has a Lagrangian for the NC system and (ii) the satisfaction of Jacobi identity is assured since Dirac brackets satisfy Jacobi identity.

  Let us now outline our formalism following   \cite{jac} {\footnote{Recent works using this framework in the context of relativistic fluids are \cite{arpan}.}}. The NC fluid model proposed by us rests essentially on the map  between the Lagrangian and Eulerian or (Hamiltonian) description of fluid dynamics. The former is based on a microscopic picture where the fluid is treated as a collection  (albeit a large number) of point particles obeying canonical Newtonian dynamics. The d.o.f.s consist of the particle coordinate and velocity, ${\bf X_i}(t), d{\bf X_i}(t)/dt$, respectively, where $i$ stands for particle index. In the limit of a continuum of particle d.o.f these reduce to  ${\bf X}({\bf x},t), d{\bf X}({\bf x},t)/dt$ with $\bf {x}$ replacing the discrete index $i$. On the other hand the Eulerian scheme starts by providing a field theory Hamiltonian $H=\int d^3r \mathcal{E} (\rho, {\bf v})$ and a set of Poisson brackets between  the fluid d.o.f that are the density and velocity fields  $\rho({\bf r},t),~{\bf v}({\bf r},t)$, respectively. The fluid equations of motion are derived from the above as Hamilton's equation of motion. There is a map that connects Lagrangian variables (${\bf X}({\bf x},t), d{\bf X}({\bf x},t)/dt$) to Eulerian fields ($\rho({\bf r},t),~{\bf v}({\bf r},t)$) and, most importantly, field theoretic Poisson algebra between Euler variables is derived from the canonical point mechanics Poisson brackets between Lagrangian variables. (These are discussed later in detail.) {\it{This is best suited for our purpose since, as discussed earlier, the NC brackets are given most naturally in point mechanics framework, that is in terms of Lagrangian variables}}. It is worthwhile to recall here that even the canonical point mechanics (Poisson) brackets lead to a quite involved and non-linear set of operatorial algebra between the Euler variables. Hence it is not entirely surprising that the simplest extension of canonical brackets to NC brackets in Lagrangian setup will lead to an involved and qualitatively distinct NC extended brackets among Euler variables. However, as we will explicitly demonstrate, {\it{these NC brackets yield a modified set of continuity equation and Euler force equation with an unchanged conserved mass and energy but NC corrected mass and momentum fluxes}}. Clearly these NC modifications will leave their marks on cosmological solutions and in particular the NC corrections will act as specific forms of cosmological perturbations.

  The paper is organized as follows: in Section 2 we recapitulate the Hamiltonian or Euler form of fluid dynamics along with its spacetime symmetries \cite{jac}. Section 3 deals with the NC fluid variable algebra  with a discussion on Jacobi identities concerning the new NC fluid algebra. Section 4 is devoted to a study of the generalized continuity and conservation principles including comments on spacetime symmetries for NC fluid. In Section 5 we provide an outline on effects on cosmological principles induced by NC modified fluid system. We conclude in Section 6 with a summary of our work and its future prospects.

\section{Canonical fluid and its symmetries in Hamiltonian framework}
Let us start with a quick recapitulation of the formulation and results for a conventional (or canonical) perfect fluid in Euler (or Hamiltonian) formulation \cite{jac}. There is a well-known prescription of starting with a microscopic or discrete picture of dynamics of a system of particles in a Lagrangian framework leading to a continuum field theory of fluid in Eulerian scheme. Infact this basic connection between the two formalisms plays the central role in the consistent generalization of the fluid in noncommutative space.

 Newton's law for the particle (Lagrangian) coordinate $X_i(t)$ and velocity $v_i(t)=\dot{X_i}$ is given by,
  \begin{equation}
 m\ddot 
  X_i(t)=m\dot v_i(t)=F_i(X(t)), \label{new}
  \end{equation}
 where $m$ is the particle mass and $F_i(X(t))$ is the applied force. The Eulerian density field for the single particle is,
  \begin{equation}
  \rho (t, {\bf r}) = m  \delta ({\bf X} (t) - {\bf r}).
  \label{rs}
  \end{equation}
 Hence the density and current fields for a collection of particles are given by, 
 
 \begin{equation}
 \rho (t, {\bf r}) = m \sum^N_{n=1} \delta ({\bf X}_n (t) - {\bf r}),
 \label{r}
 \end{equation}
 
 \begin{equation}
 {\bf j} (t, {\bf r}) = {\bf v} (t, {\bf r}) \rho (t, {\bf r}) = m \sum^N_{n=1} {\bf \dot{X}}_n (t) \delta ({\bf X}_n (t) - {\bf r}).
 \label{j}
 \end{equation}
 For fluid field theory the discrete particle labels are replaced by continuous variables and we end up with,
 \begin{eqnarray}
  \rho({\bf r})=\rho_0\int\delta(X(x)-r)dx,\label{ncc14} \\
  v_i({\bf r})=\frac{\int dx \dot{X_i}(x)\delta (X(x)-r)}{\int dx \delta(X(x)-r)}.
 \label{ncc13}
 \end{eqnarray}

 From now on we will omit  $t$ from the arguments of the d.o.f.. In a Hamiltonian formulation the canonical  Poisson bracket structure  is given by
  \begin{equation}
 [\dot{X}^i , X^j ] = (i/m) \delta^{ij},~~ [X^i , X^j ] = 0,~~ [\dot{X}^i , \dot{X}^j ] = 0.
 \label{nnp1}
 \end{equation}
 For the Lagrangian fluid this is generalized to   \cite{jac}, 
 \begin{equation}
 \{\dot{X}^i ({\bf x}), X^j ({\bf x'})\} = \frac{1}{\rho_0} \delta^{ij} \delta ({\bf x} -{\bf x'});~~ \{X^i ({\bf x}), X^j ({\bf x'})\} =  \{\dot{X}^i ({\bf x}), \dot{X}^j ({\bf x'})\} = 0.
 \label{p11}
 \end{equation}
 Obviously the above bracket structure satisfies the Jacobi identity.
Using the definitions of $\rho$ and $\bf j$ in terms of $\bf X$ and $\bf \dot{X}$ given above,  a straightforward but somewhat non-trivial computation leads to the Poisson algebra between the Euler variables $\rho$ and ${\bf j}$: 
\begin{eqnarray}
\{\rho({\bf r}), \rho ({\bf r}')\} &=&0 \label{1rj},\\
\{j^i ({\bf r}), \rho ({\bf r'})\}&=& \rho ({\bf r}) \partial_i \delta \brr \label{2rj},\\
\{j^i ({\bf r}),j^j ({\bf r}')\} &=& j^j ({\bf r}) \partial_i \delta \brr + j^i ({\bf r'}) \partial_j
\delta \brr . \label{rj}
\end{eqnarray}
Since ${\bf j}={\bf v} \rho$ an equivalent set of brackets follow:
\begin{eqnarray}
\{v^i({\bf r}), \rho({\bf r'})\} = \partial_i \delta \brr, \qquad \label{rrv}\\
\{v^i({\bf r}), v^j({\bf r'})\} = -\frac{\omega_{ij} ({\bf r})}{\rho({\bf r})} \delta \brr,
\label{rv}
\end{eqnarray}
where
\begin{equation}
\omega_{ij} ({\bf r}) = \partial_i \ v_j {(\bf r)}-\partial_j v_i ({\bf r})
\label{v}
\end{equation}
is called the fluid vorticity. 

However, for $\omega_{ij}=0$, that will be considered later, the current algebra (\ref{2rj}-\ref{rj}) changes slightly to
\begin{eqnarray}
\{\rho({\bf r}), \rho ({\bf r}')\} &=&0 \label{1jj},\\
\{j^i ({\bf r}), \rho ({\bf r'})\}&=& \rho ({\bf r}) \partial_i \delta \brr \label{2jj},\\
\{j^i ({\bf r}),j^j ({\bf r}')\} &=& j^j ({\bf r}) \partial_i \delta \brr + j^i ({\bf r'}) \partial_j
\delta \brr +(\partial_j j^i({\bf r})-\partial_ij^j({\bf r})) \delta \brr . \label{jj}
\end{eqnarray}

Positing the Hamiltonian for a barotropic fluid with potential energy $U(\rho )$ as
\begin{equation}
H = \int d  r \bigg(\frac{1}{2} \rho  {\bf v}^2 + U(\rho) \bigg),
\label{hc}
\end{equation}
 the equations of motion turn out to be
\begin{eqnarray}
\dot{\rho} = \{H, \rho\} = -\vec \nabla \cdot ({\bf v} \rho )
\label{E},\\
\dot{\bf v} = \{H, {\bf v}\} = - ({\bf v} \cdot \vec \nabla) {\bf v} -
\vec \nabla U' (\rho).
\label{e} 
\end{eqnarray}
The above equations (\ref{E}), (\ref{e}) constitute respectively the continuity equation and Euler (force) equation, the two central equations governing perfect fluid dynamics.

Furthermore the Hamiltonian formalism is very convenient for studying the symmetries of the system. Renaming the densities in a suggestive way for a generalization to relativistic fluid \cite{jac}, (which will not be pursued here),
\begin{equation}
\mathcal{E} = \frac{1}{2} \rho {\bf v}^2 + U = T^{00},
\label{en}
\end{equation}
together with the energy flux
\begin{equation}
T^{jo} = \rho v^j (\frac{1}{2} {\bf v}^2 +  U'),
\label{ef}
\end{equation}
one derives the energy conservation law,
\begin{equation}
\dot{T}^{oo} + \partial_j T^{jo} = 0 . 
\label{ec}
\end{equation}
 The momentum density, $\cal P$, 
\begin{equation}
		{\cal P}^i = \rho v^i  = T^{oi},
		\label{p1}
\end{equation}
	and the stress tensor $T^{ij}$ with $P=\rho U' - U$ defined as the pressure,
	\begin{equation}
		T^{ij} = \delta^{ij} (\rho U' - U) + \rho v^i v^j = \delta^{ij} P +  \rho v^i v^j,
		\label{t1}
	\end{equation}
satisfy
\begin{equation}
	\dot{T}^{oi} + \partial_j T^{ji} = 0 . 
	\label{t2}
\end{equation}
The time translation and space translation invariances are assured by the conserved quantities,
\begin{equation}
E = \int d  x \mathcal{E} \qquad \mbox{(time-translation)},
\label{t}
\end{equation}
\begin{equation}
{\bf P} = \int d r \ \vec{\cal P} =  \int d r ~ {\bf j} \qquad
\mbox{(space-translation)}.
\label{s}
\end{equation}
 Rotational invariance is preserved by the conservation of  angular
momentum,
\begin{equation}
M^{ij} = \int d  r\ (r^i \mathcal{P}^j - r^j \mathcal{P}^i) \qquad \mbox{(spatial rotation)}.
\label{am}
\end{equation}
Furthermore invariance of the non-relativistic theory under  Galilean transformation   yields the Galileo
boost constant of motion,
\begin{equation}
{\bf B} = t\ {\bf P} - \int d r\ {\bf r} \rho \qquad \mbox{(velocity boost)}.
\label{g}
\end{equation}
The continuity equation provides the total number or mass as the final conserved quantity
\begin{equation}
N = \int dr \rho \qquad \mbox{(number)}.
\label{n}
\end{equation}
It is quite natural in fluid dynamics to consider irrotational fluids as a special case so that  $\omega_{ij} =0$ and we will also adopt this choice for simplicity but as well as for more serious reasons that will become clear as we proceed. We will compare and contrast these important features of the ideal fluid with  the NC generalized fluid model that will be formulated in the next section.\\

\section{Noncommutative extension of fluid variable brackets}
In NC quantum mechanics it is customary to consider NC space as
\begin{equation}
[\dot{X}^i , X^j ] = (i/m) \delta^{ij},~~ [X^i , X^j ] = i\theta ^{ij},~~ [\dot{X}^i , \dot{X}^j ] = 0.
\label{nnp}
\end{equation}
Note that this is the simplest extension that describes an NC space (for applications see \cite{bertol}). Obviously, as explained in the Introduction, this and similar extensions by non-operatorial terms, such as $\theta_{ij}$ here, satisfies Jacobi identity and hecnce are examples of perfectly consistent bracket structure. 
where the antisymmetric  NC parameter $\theta_{ij}=-\theta_{ji}$ is  constant. From High Energy Physics perspective bounds on $\theta $, ($\theta \le (10 TeV)^{-2}$), are available \cite{theta}. Although this is very small keeping in mind present day experimental results its effect might turn out to be  significant in a classical macroscopic system such as a fluid. {\footnote{ Indeed it is possible to consider more general forms of noncommutativity with additional constant terms in rest of the brackets or even consider operatorial forms of NC structures such as Kappa-Minkowski algebra or brackets compatible with Generalized Uncertainty Principle \cite{gup}. NC fluid dynamics, compatible with a Snyder form \cite{sny} of operatorial spacetime has been discussed in \cite{van}.  However, for the type of applications we have in mind,  the simplest NC extension studied here induces very interesting and non-trivial effects in the fluid model and so other alternatives have not been considered, although these can be studied in our approach.}}

As advertised in the Introduction and following (\ref{p11}) \cite{jac}, we introduce the NC algebra among the  Lagrangian degrees of freedom  as
\begin{equation}
\{\dot{X}^i ({\bf x}), X^j ({\bf x'})\} = \frac{1}{\rho_0} \delta^{ij} \delta ({\bf x} -{\bf x'}),~~ \{X^i ({\bf x}), X^j ({\bf x'})\} =\frac{1}{\rho_0} \theta ^{ij}\delta ({\bf x} -{\bf x'}),~~ \{\dot{X}^i ({\bf x}), \dot{X}^j ({\bf x'})\} = 0.
\label{p}
\end{equation}
Once again this NC algebra satisfies Jacobi identity and is a perfectly consistent example of bracket structure.
This is a classical analogue of (\ref{nnp}) applied to Lagrangian coordinates and a NC generalization of (\ref{p11}). This simple form of NC extension and its subsequent generalization to NC quantum field theory started after the seminal work of Seiberg and Witten \cite{sw} who showed its relevance in  certain low energy limits of  string theory where the open string end points reside on NC $D$-branes and $\theta ^{\mu\nu}$ gets identified with the anti-symmetric field $B^{\mu\nu}$.  In the same way as was done for the normal fluid \cite{jac}, the NC algebra (\ref{p}) generates the  NC Euler variable algebra. A form of "exotic plasma" fluid where the particle Lagrangian coordinates are endowed with an "exotic" algebra, (that is effectively a NC system), was studied in \cite{hor}. In another approach \cite{van} NC effects, (of the same canonical form used here), were incorporated directly in the form of star products in the fluid action in Ka¨hler parametrization to NC spacetimes.  For completeness we show a few steps explicitly. Using the defining equations for Euler variables from (\ref{ncc14}, \ref{ncc13}), we need to compute,
\begin{eqnarray}
\{ \rho ({\bf r}),\rho ({\bf  r'})\}=\rho_0^2\left[\int \delta(X(x)-r)dx,\int \delta(X(y)-r')dy\right],
\label{nc9}
\end{eqnarray}
\begin{eqnarray}
\{v_i({\bf r}),\rho({\bf r'})\}=\left[\frac{\int dx \dot{X_i}(x)\delta (X(x)-r)}{\int dx \delta(X(x)-r)},\rho_0\int\delta(X(y)-r')dy\right],
\label{nc13}
\end{eqnarray}
\begin{eqnarray}
\{v_i({\bf r}),v_j({\bf r'})\}=\left[\frac{\int dx \dot{X_i}(x)\delta (X(x)-r)}{\int dx \delta(X(x)-r)},\frac{\int dy \dot{X_j}(y)\delta (X(y)-r')}{\int dy \delta(X(y)-r')}\right],
\label{nc8}
\end{eqnarray}
where the NC algebra (\ref{p}) between the Lagrange particle coordinates have to be used. We obtain,
\begin{eqnarray}
\{ \rho ({\bf r}),\rho ({\bf  r'})  \}=\theta_{ij}\partial_i\rho({\bf r})\partial_j \delta \brr ,
\label{e2}
\end{eqnarray}
\begin{eqnarray}
\{ v_i({\bf r}),\rho ({\bf  r'})  \}=\partial _i \delta \brr -\theta_{jk}\partial _j v_i({\bf r}) \partial_k \delta \brr ,
\label{e1}
\end{eqnarray}
\begin{eqnarray}
	~~~~~~~~~~~~ \{v_i({\bf r}),v_j({\bf r'})\}&=&\frac{(\partial_jv_i-\partial_iv_j)}{\rho}\delta \brr \nonumber \\
	&&+\theta_{kl}[\partial_l\delta \brr(\frac{\partial_k(v_iv_j)}{\rho}-v_iv_j\partial_k(\frac{1}{\rho}))+(\frac{\partial_kv_i\partial_lv_j}{\rho}-\partial_l(v_iv_j)\partial_k(\frac{1}{\rho}))\delta \brr] \nonumber \\
	&&-\frac{\rho_0\theta_{kl}}{\rho({\bf r})\rho({\bf r'})}\partial_l\delta \brr \frac{\partial}{\partial r_k}[\int dx \dot{X_i}(x)\dot{X_j}(x)\delta(X(x)-r)] .
	\label{nc7}
\end{eqnarray}
Comparing with (\ref{rrv}-\ref{rv}) we immediately notice that a rich NC algebra has emerged due to the mapping between Lagrange and Euler degrees of freedom, which reduces to the canonical form (\ref{rrv}, \ref{rv}) for $\theta^{ij}=0$. However we also encounter a closure problem (that is typical of non-linear systems) in that the $\{v_i, v_ j\} $ bracket contains higher moments of $v_i$. First of all let us consider irrotational fluid (in the canonical case that is with $\theta_{ij} =0$) having zero vorticity $\omega_{ij} = \partial_iv_j-\partial_jv_i =0$, which gets rid of the canonical part of the $\{v_i, v_ j\} $ bracket. This simplification is a common practice in various contexts of fluid dynamics. But, interestingly enough, even with this restriction, $O(\theta )$ terms in $\{v_i({\bf r}),v_j({\bf r'})\}$ survive showing that a noncommutativity induced vorticity has reappeared. This is reminiscent of our previous results that NC can induce spin in mechanical systems \cite{sg}.  Hence to make life simpler, at least for the present work, we will further assume that  $O(\theta v^2/\rho )$ terms are small and omit them, indicating that we are working in low velocity (or energy) and high density regime. Throughout our work we will exploit the relation $\omega_{ij} = \partial_iv_j-\partial_jv_i =0$ many times.

 Hence after all these restrictions of irrotational fluid at low velocity and high density we have been able to provide an extended fluid algebra which has been derived from basic principles and has its origin in conventional NC point mechanics. The cherished NC fluid algebra,  
\begin{equation}
\{ v_i({\bf r}),\rho ({\bf  r'})  \}=\partial _i \delta \brr -\theta_{jk}\partial _j v_i({\bf r}) \partial_k  \delta \brr ,
\label{e11}
\end{equation}
\begin{equation}
\{ \rho ({\bf r}),\rho ({\bf  r'}) \}=\theta_{ij}\partial_i\rho({\bf r})\partial_j \delta \brr ,
\label{e12}
\end{equation}
\begin{equation}
\{v_i({\bf r}), v_ j ({\bf  r'})  \}=0,
\label{e13}
\end{equation}
constitute our primary major result. Incidentally, it should be pointed out that a structure similar to  (\ref{e12}) appeared earlier  in \cite{gural} (although the full algebra (\ref{e11}-\ref{e13}) was not provided) and furthermore it emerged in a totally different framework in magnetohydrodynamics (as discussed in the Introduction) and was applied in a different context.  Rest of the paper deals with studying some of the fluid properties compatible with this NC algebra.

Once again it is straightforward to recover the NC current algebra, 
\begin{eqnarray}
\{ \rho ({\bf r}),\rho ({\bf  r'}) \}&=&\theta_{kl}\partial_k\rho({\bf r})\partial_l \delta \brr ,  \label{n1j}\\
\{ j_i ({\bf r}),\rho ({\bf  r'}) \}&=&\rho({\bf r})\partial_i \delta \brr +\theta_{jk}[2j_i\frac{\partial_j\rho}{\rho}-\partial_j j_i]\mid_r \partial_k\delta\brr ,\label{2nj}  \\
\{ j_i ({\bf r}),j_j ({\bf  r'}) \}&=&\left[\left(j_i(\delta_{jl}-\theta_{kl}\partial_k\frac{j_j}{\rho})\right)\mid_r+\left(j_j(\delta_{il}-\theta_{kl}\partial_k\frac{j_i}{\rho})\right)\mid_{r'}\right]\partial_l\delta\brr \nonumber \\
&&+\left[(\partial_jj_i-\partial_ij_j)-\theta_{kl}\partial_l(\frac{1}{\rho})(j_j\partial_kj_i-j_i\partial_kj_j)\right]\delta\brr   \label{3nj}
\end{eqnarray}
where the subscripts $r$ and $r'$ denote that the arguments of the functions in t parenthesis   are at  $r$ and $r'$ respectively.

However, before proceeding further with a new Poisson-like structure, (especially if it is non-linear and operatorial in nature), it is imperative to ensure that it satisfies the Jacobi identities. \\
{\textbf {Jacobi identity considerations}}: Jacobi identity plays a vital role in the internal consistency of the commutator structure in quantum mechanics or quantum field theory. Although, in terms of commutators, it appears as a trivial identity,
\begin{equation}
	J(A,B,C)\equiv 	[[A,B],C]+[[B,C],A]+[[C,A],B] =0
	\label{1}
\end{equation}
 leading to compatibility with Jacobi identity, actual situation is much more tricky in quantum field theory due to clash between singularities and limiting procedures and this requirement yields non-trivial consequences (see for example \cite{jacobi}). Its classical analogue, defined in terms of Poisson brackets (or Dirac brackets in constrained systems)
\begin{equation}
J(A,B,C)\equiv 	\{\{A,B\},C\}+\{\{B,C\},A\}+\{\{C,A\},B\}=0
\label{1}
\end{equation}
imposes an equally important restriction on the brackets. Indeed, for canonical brackets as in (\ref{1}) as well as its extension by $c$-number functions as in (\ref{1}) or more generalized versions, the extended NC brackets trivially satisfy $J(A,B,C)=0$ without imposing any restrictions on the extensions. The double commutators will vanish trivially so long as the single commutators  do not involve operators on the RHS. However the situation changes drastically once the brackets involve operators, such as the brackets among the fluid variables given in (\ref{e11}-\ref{e13}). Naively one might think that if the canonical algebra (\ref{1rj}-\ref{rj}) satisfied Jacobi identities, (which it does \cite{jac}), the present set should also satisfy  Jacobi identities since after all only constant NC extensions have been taken into account in the parent Lagrangian coordinates (\ref{p}). However, checking that the algebra (\ref{e11}-\ref{e13}) satisfies Jacobi identities is somewhat non-trivial as we briefly demonstrate below.

For field theory one has the requirement that Jacobi identities have to be obeyed {\it{locally}}:
\begin{equation}
		J(A(x),B(y),C(z))\equiv 	\{\{A(x),B(y)\},C(z)\}+\{\{B(y),C(z)\},A(x)\}+\{\{C(z),A(x)\},B(y)\}=0.
	\label{2}
\end{equation}
Writing in detail for the density variable $\rho (x)$,
\begin{eqnarray}
J\{\rho,\rho,\rho\}\equiv\{\{\rho(x),\rho(y)\},\rho(z)\}+cyclic ~~terms,
\label{nc5}
\end{eqnarray}
we find
\begin{eqnarray}
\{\{\rho(x),\rho(y)\},\rho(z)\}=-\theta^{ij}\theta^{kl}\partial_k\rho(z)\partial_j^y\partial_i^z\partial_l^z(\delta(x-y)\delta(y-z))=0,
\end{eqnarray}
which vanishes due to  support of  RHS only for $\bf x=\bf y=\bf z$ and anti-symmetry of $\theta ^{ij}$, 
 thus leading to 
\begin{eqnarray}
J\{\rho,\rho,\rho\}\equiv\{\{\rho(x),\rho(y)\},\rho(z)\}+cyclic~~terms=0.
\label{nc5}
\end{eqnarray}
In a similar way for the mixed  $J\{\rho (x),\rho (y),v_i(z)\}$ we have
\begin{eqnarray}
\{\{\rho(x),\rho(y)\},v_i(z)\}=-\theta^{kl}\left[\partial_k^y\partial_l^z\partial_i^z(\delta(x-y)\delta(x-z))+\theta^{mn}\partial_kv_i(z)\partial_n^y\partial_k^z\partial_l^z(\delta(x-y)\delta(x-z))\right]=0
\end{eqnarray}
due to  support of  RHS only for $\bf x=\bf y=\bf z$ and anti-symmetry of $\theta ^{ij}$. The other two cyclic terms also vanish for the same reason yielding,
\begin{eqnarray}
J\{\rho,\rho,v_i\}\equiv\{\{\rho(x),\rho(y)\},v_i(z)\}+cyclic~~terms=0.
\label{nc6}
\end{eqnarray}
Rest of the Jacobi identities $J(\rho ,v_j ,v_i)$ and $J(v_k ,v_j ,v_i)$ are trivially satisfied. Thus the new algebra (\ref{e11}-\ref{e13}) provided by us proves to be a consistent extension of irrotational fluid dynamics algebra in noncommutative space. 

\section{Properties of NC fluid: conservation laws and symmetries}
Let us begin by discussing the symmetries and conservation laws of the NC fluid as a Hamiltonian system.\\
{\textbf {Conservation laws}}: As a trial form we assume  same expression for the Hamiltonian as the canonical one (\ref{hc}). Exploiting the NC algebra (\ref{e11},\ref{e12},\ref{e13}) we
compute the NC fluid equations of motion:
\begin{eqnarray}
\dot{\rho}=  \{H, \rho\} =  -\partial_i(\rho v_i+\theta_{ij}\rho \partial_j v^2),
\label{nc1}
\end{eqnarray}
\begin{eqnarray}
\dot{v_i}=  \{H,  v_i\} =   -\partial_i(\frac{v^2}{2}+U')+\theta_{jk}(\partial_i v_j)\partial_k(\frac{v^2}{2}+U')\\
=-\partial_j\left((\frac{v^2}{2}+U')\delta^{ij}+\theta_{jk}(\frac{v^2}{2}+U')(\partial_k v_i)\right ).
\label{nc2}
\end{eqnarray}
We stress that in the NC generalized $\rho$-equation of motion (\ref{nc1}) can still be interpreted as a conservation law since the NC correction term also is a total derivative, leading to 
 the  conservation of total number $N$ as defined in (\ref{n})  in the NC generalization. {\footnote{It is tempting to redefine the velocity field $v_i \rightarrow \bar{v_i}=v_i+\theta_{ij} \partial_j v^2$ so that (\ref{nc1}) can be recast as $\dot \rho +\partial_i (\rho \bar{v_i})$ but unfortunately this does not lead to a simpler bracket structure with $\bar{v_i},\rho $ and so we do not consider it.}} In the NC modified Euler equation out of the $\theta$-terms, $(\theta_{jk}\partial_i v_j)\partial_k(\frac{v^2}{2})$ can be dropped in the approximation we have considered but  $(\theta_{jk}\partial_i v_j)\partial_kU'$ will survive. Later on, in the context of cosmology, we will demonstrate that the $\theta$-terms in these two basic fluid equations provide the seeds for cosmological perturbations.

Next we come to the all important equation that deals with energy conservation. Exploiting the equations for $\dot \rho $ and $\dot {v_i}$  from (\ref{nc1}, \ref{nc2}) and using 
\begin{eqnarray}
\mathcal{E}=\frac{\rho v^2}{2}+U(\rho)
\end{eqnarray}
we obtain
\begin{eqnarray}
\dot{\mathcal{E}}=-\partial_i\left[\rho v_i(\frac{v^2}{2}+U')+\theta_{ij}\{-\partial_jv^2(\rho\frac{v^2}{2}+U)+\frac{v^2}{2}\partial_jP\}\right],
\label{nc3}
\end{eqnarray}
where  $P=\rho U'-U$ is the pressure. Very interestingly the energy conservation law is maintained since, once again as it happened for the number conservation law in (\ref{nc1}), the extra term is a total derivative. Thus the canonical form of energy density is preserved but going back to our earlier identification (\ref{ef}) we find that the energy flux $T^{j0}$ receives an NC correction,
\begin{equation}
T^{j0}=\rho v_i(\frac{v^2}{2}+U')+\theta_{ij}\{-\partial_jv^2(\rho\frac{v^2}{2}+U)+\frac{v^2}{2}\partial_j(\rho U'-U)\}.
\label{nct}
\end{equation}
{\textbf {Space-time symmetries}}: Let us now check translation invariance of the NC fluid system. Keeping the expression for  momentum $\pi ^i$ unchanged we define $\Pi^i =\int dr \pi^i$ and compute,
\begin{equation}
\{\Pi^i,\rho ({\bf  r'})  \}=-\partial_i \rho ,
\label{rr}
\end{equation}
\begin{equation}
\{\Pi^i,,v_j ({\bf  r'})  \}=-\partial_i v_j+\theta_{kl}\partial_k v_j\partial_l v_i.
\label{vv}
\end{equation}
Notice that $\bf{\Pi }$ translates $\rho $ correctly in the NC space but fails to do so for $\bf{v}$. However, pursuing with definition for $\pi_i$ we we compute its time derivative to find
\begin{eqnarray}
\pi^i=\rho v_i ,~~ \dot{\pi^i}=\partial_j\left[-\{\rho v_iv_j+\delta_{ij}P\}+\theta_{jk}\{v_i\partial_kP-v_i\rho\partial_kv^2\}\right]+\frac{3}{2}\theta_{jk}\rho(\partial_jv_i)\partial_kv^2.
\label{nc4}
\end{eqnarray}
It is worthwhile to comment on this relation. First of all note that the last term in RHS, not being a total derivative, apparently breaks the total momentum conservation of  $\Pi^i $ but we can safely drop the disturbing last term in our approximation scheme, leading to a slightly "weaker" momentum conservation principle. But this is to  be expected since $\theta$-correction acts as an external force as we have seen in the modified Euler equation (\ref{nc2}).

Thus we end up with 
\begin{equation}
\dot{\pi^i}=\partial_j\left(-(\rho v_iv_j+\delta_{ij}P)+\theta_{jk}v_i\partial_kP\right), 
\label{tij}
\end{equation}
and define a modified $T^{ij}$,
\begin{equation}
T^{ij}=-(\rho v_iv_j+\delta_{ij}P)+\theta_{jk}v_i\partial_kP.
\label{tt}
\end{equation}
We immediately observe that the $\theta$-contribution in $T^{ij}$ is not symmetric under interchange of $i,j$. Let us return to the conservation of angular momentum $M^{ij}$ in the canonical fluid ($\theta^{ij}=0$). This requires using the equations (\ref{p1},\ref{t1},\ref{t2},\ref{am}) together with the fact that (\ref{t1}) is symmetric. Since  the NC correction in $T^{ij}$ is not symmetric the total angular momentum will not be conserved for the NC fluid and in particular
\begin{eqnarray}
\dot{M_{ij}}&=&\int dr (r^i \dot{\pi^j}-r^j\dot{\pi^i}) \nonumber \\
&&=\int dr \theta_{kl}[r^i\partial_jv_l-r^j\partial_iv_l](\partial_kv^2) \rho .
\label{nc12}
\end{eqnarray}
Again this result is also expected since we have introduced a constant set of parameters $\theta^{ij}$ that do not transform at all under rotations so that the fundamental NC bracket in (\ref{p}), $\{X^i ({\bf x}), X^j ({\bf x'})\} = \frac{\theta_{ij}}{\rho_o}\delta ({\bf x} -{\bf x'})$ fails to transform covariantly under spatial rotation. On the other hand the Galileo
boost constant of motion $\bf{B}$ (\ref{g}) will be preserved in the NC space in our approximation scheme.
\section{Cosmological implications} In this section we wish to outline how the NC extended fluid equations, in particular the NC Euler equation (\ref{nc2}), can introduce density perturbations in cosmology (for basic cosmology see for example \cite{lid}). There are several NC extensions of cosmological models, such as \cite{nccos1} (where the NC geometry  is imposed from the spectral point of view \cite{con}). On the other hand, in  \cite{nccos2} noncommutativity is introduced from spacetime uncertainty relation and the formalism requires spacetime noncommutativity in an essential way. (For question related to unitarity of models with spacetime noncommutativity see comment and references in \cite{nccos2}.) These models are very different from our way of introducing NC at Lagrangian coordinate level, explained earlier.

Before presenting our formulation of NC gravity-fluid system let us emphasize that we are considering a simplified system where gravity appears in a mean-field approximation. This is because the NC extension of a fully interacting gravity-fluid system is very complicated and we restrict ourselves to a scenario where gravity is external in the sense that it is not affected by the NC modification of the fluid.

However, before proceeding further, it is crucial to understand how the $\theta_{\mu\nu} $ form of NC can (and if at all at least for the present work) affect the pure (Einstein) gravitational sector. In this context, (as mentioned in the Introduction), we need to exploit the other alternative path leading to NC-extension, that is   a  direct application of  the Groenwald-Moyal $*$-product \cite{sw,ncrev} in Einstein General Theory action to obtain the NC-extended  field theory action. Now, gravity can be considered as a gauge theory and \cite{sw} provides a prescription to generalize a gauge theory in NC space via the Seiberg-Witten map \cite{sw}. But the problem in this naive application is that the fundamental NC structure  (\ref{nnp}) is not  invariant under general
 coordinate transformation. It was first noticed by Calmet and Kobakidze \cite{ck1} that there exists a  subclass of general coordinate transformations that are compatible with (\ref{nnp}) leading to a version of  General
 Relativity based on volume-preserving diffeomorphism. Interestingly, this was already  known as 
  the unimodular theory of gravitation \cite{umod} since the Jacobian of this restricted coordinate transformation is
  equal to $1$, meaning that the volume  is invariant, (although space-time volume may not be preserved).

   This model with the gauge symmetries intact was generalized to NC spacetime (via $*$-product and Seiberg-Witten map) but it turned out that due to symmetry requirements the $O(\theta)$ corrections miraculously cancel and hence to first order in $\theta$ Einstein action and its NC extension are identical \cite{pm}.  Thus we come to the
   conclusion that symmetries of canonical noncommutative
   spacetime naturally lead to the noncommutative version of
   unimodular gravity for $O(\theta )$ results.  Furthermore, there exists other distinct scenarios of NC generalizations, such as discussed in \cite{other} and in all the cases the $O(\theta )$ contribution is absent. Incidentally, the non-zero NC corrections at $O(\theta ^2)$ were computed in \cite{ck2}. In this regard the NC corrections in the matter (fluid) sector presented in our work are linear in $\theta $ and hence we are allowed to work with the cosmological model where the gravity sector remains unchanged and NC modifications are manifested only in the matter sector.

 As is customary in cosmological context one works in a comoving frame and the starting equations in Friedmann-Robertson-Walker (FRW) framework of cosmology are the continuity and Euler equations,
\begin{eqnarray}
\dot{\rho}=-3H(\rho+P)=-3\frac{\dot{a}}{a}(\rho+P),
\label{nc14}
\end{eqnarray}
\begin{eqnarray}
\frac{\ddot{a}}{a}=-\frac{\rho+3P}{6M^2}+\frac{\Lambda}{3}.
\label{nc15}
\end{eqnarray}
$P,\Lambda $ denote the pressure and Cosmological constant respectively and $a(t)$ is the scale factor. $M=(8\pi G)^{-1/2}$ refers to Newton's constant $G$. Introducing the Hubble parameter $H(t)={\dot a}/a$ we recover the  Friedmann equation,
\begin{eqnarray}
\frac{\dot{a}^2}{a^2}=H^2=\frac{\rho}{3M^2}+\frac{\Lambda}{3}-\frac{k}{a^2}.
\label{nc16}
\end{eqnarray}
Let us now connect the above with the fluid dynamics discussed so far. We need to cast the dynamical fluid equations in  comoving coordinates defined as
\begin{equation}
{\bf r} =a(t){\bf x}(t),
\label{c1}
\end{equation}
where ${\bf r}(t), {\bf x}(t)$ and $a(t)$ denote respectively the proper coordinates, comoving coordinates and the  scale factor exhibiting the expansion of the universe. The  ${\bf r}(t)$ coincides with our (fluid) position coordinate. The general relation between proper and comoving velocities is given by 
\begin{equation}
\dot{\bf r} =H(t){\bf r}+a\dot {\bf x}(t).
\label{cc2}
\end{equation}
Hence the "peculiar velocity" $\dot {\bf x}(t)$ signifies deviation from the perfect Hubble flow.
In conventional cosmological FRW model one  considers the comoving coordinates ${\bf x}(t)$ to be time independent so that the comoving velocity $\dot {\bf x}(t)$ is zero.    In our notation (\ref{cc2}) will be
\begin{equation}
{\bf v} =H(t){\bf r} +{\bf u},
\label{c2}
\end{equation}
with ${\bf u}$ defined as the peculiar velocity. We rewrite the NC fluid equations once again,
\begin{eqnarray}
\frac{\partial \rho}{\partial t}|_r +\frac{\partial}{\partial r_i}\left(\rho v_i+\theta_{ij}\rho \partial_j v^2\right)=0.
\label{c1}
\end{eqnarray}

\begin{eqnarray}
\frac{\partial v_i}{\partial t}|_r+v_j\partial_j v_i=-\frac{\partial _i P}{\rho}+\frac{\theta_{jk}\partial_iv_j \partial_k P}{\rho}-\partial_i \varPhi
\label{c2}
\end{eqnarray} 
where, we introduce $\varPhi$ as a generic potential. We need to recast the dynamics in the comoving coordinates ${\bf {x}},t$. The space derivatives are easily related by
$$\partial /\partial {\bf{r}} =(1/a)\partial /\partial {\bf{x}}.$$ On the other hand, the time derivatives at constant ${\bf {r}}$ and constant  ${\bf {x}}$ are related by,
$$\frac{\partial}{\partial t}\mid_{\bf{r}}=\frac{\partial}{\partial t}\mid_{\bf{x}}-\frac{\dot a}{a}({\bf{x}}.{\partial_{\bf{x}}}).$$ 
Furthermore note that the "no vorticity condition" $\omega_{ij}=0$ in comoving coordinates is expressed as
\begin{eqnarray}
\frac{1}{a}\frac{\partial}{\partial x_i}(\dot{a}x_j+u_j)=\frac{1}{a}\frac{\partial}{\partial x_j}(\dot{a}x_i+u_i) .
\label{c3}
\end{eqnarray}
Simplifying the above to
\begin{eqnarray}
\dot{a}\delta_{ij}+\frac{\partial u_j}{\partial x_i}=\dot{a}\delta_{ij}+\frac{\partial u_i}{\partial x_j}
\label{c4}
\end{eqnarray}
we find
\begin{eqnarray}
\frac{\partial u_j}{\partial x_i}=\frac{\partial u_i}{\partial x_j}.
\label{c5}
\end{eqnarray}
This relation will be used subsequently. Exploiting the above identities we derive the cherished expressions for the NC fluid dynamics in comoving frames:
\begin{eqnarray}
\frac{\partial \rho}{\partial t}|_x+\frac{1}{a}\frac{\partial}{\partial x_i}(\rho u_i)+\frac{3\rho \dot{a}}{a}+\frac{2\theta_{ij}}{a^2}\frac{\partial\rho}{\partial x_i}(\dot{a}^2x_j+\dot{a}u_j+(\dot{a}x_k+u_k)\frac{\partial u_j}{\partial x_k})=0,
\label{c6}
\end{eqnarray}
\begin{eqnarray}
\frac{\partial u_i}{\partial t}+\frac{\dot{a}}{a}u_i+\frac{u_j}{a}\frac{\partial u_i}{\partial x_j}=-\frac{1}{a\rho}\frac{\partial P}{\partial x_i}-\frac{1}{a}\frac{\partial \phi_{pec}}{\partial x_i}+ \theta_{ik}\frac{\dot{a}}{\rho a^2}\frac{\partial P}{\partial x_k}+\theta_{jk}\frac{1}{\rho a^2}\frac{\partial u_j}{\partial x_i}\frac{\partial P}{\partial x_k}-(\ddot{a}+\frac{4\pi}{3}aG\rho_b)x_i .
\label{c7}
\end{eqnarray}
Without the peculiar velocity (${\bf{u}}$-dependent) terms the NC continuity equation (\ref{c6}) reduces to 
\begin{eqnarray}
\dot{\rho}+\frac{3\rho\dot{a}}{a}+\psi=0,
\label{c9}
\end{eqnarray}
where $\psi=\frac{2\theta_{ij}}{a^2}\frac{\partial \rho}{\partial x_i}\dot{a}^2 x_j$ is the NC correction. {\footnote{As an aside we note that the conventional continuity equation for fluid can be recovered from first law of thermodynamics for adiabetic processes. Hence the NC correction brings in some sort of non-adiabeticity in the process.}} On the other hand, the ${\bf{x}}$-dependent term in (\ref{c7}) yields
\begin{eqnarray}
\ddot{a}+\frac{4\pi}{3}aG\rho_b =0,
\label{cc9}
\end{eqnarray}
which agrees with (\ref{nc15}) provided $\Lambda =0$ and $P=0$ and  a spherically symmetric dust potential is considered, that is
\begin{eqnarray}
\Phi = (2/3)\pi G\rho_b(a{\bf{x}})^2 +\phi_{pec}.
\label{cc}
\end{eqnarray}
Here $\phi_{pec}$ constitutes the peculiar potential. Indeed, for a barotropic fluid whose potential $U(\rho )$ depends on the density $\rho $ only and furthermore specializing to a dust,  with $U(\rho )\sim \rho$, its' pressure vanishes, $P=\rho U'-U=0$. (These restrictions can be relaxed in a straightforward way.)

Let us note the consequences of an NC space. From (\ref{c9},\ref{cc9}) it is clear that to our level of approximation, even the so-called "unperturbed" universe with ${\bf{u}}=0$ receives NC contributions. From (\ref{c9},\ref{cc9}) we recover
\begin{eqnarray}
\frac{1}{2}\frac{d}{dt}(\dot{a}^2)=\frac{4\pi G}{3}\frac{d}{dt}(\rho a^2)+\frac{4\pi G}{3}\psi a^2 ,
\label{nc19}
\end{eqnarray}
leading to  a modified Friedmann equation,
\begin{eqnarray}
\frac{\dot{a}^2}{a^2}=\frac{8\pi G \rho}{3}-\frac{k}{a^2}+\frac{8\pi G }{3}\frac{1}{a^2}\int a^2 \psi dt,
\label{c10}
\end{eqnarray}
where $k$ is a constant. Hence for non-zero pressure we see that the ${\bf{u}}$-equation of motion from the Euler equation (\ref{c7}) will be modified by $\theta $-contributions.  We do not wish to dwell too much on NC effects on cosmological (density) perturbations in the present paper as these will be analyzed in a separate publication. But just to indicate a straightforward NC effect, let us express (\ref{c10}) as
\begin{eqnarray}
\frac{\dot{a}^2}{a^2}=\frac{8\pi G \rho}{3}-\frac{k_{nc}}{a^2},
\label{c100}
\end{eqnarray}
where $k_{nc} = k-\frac{8\pi G}{3}\int a^2 \psi dt$. Conventionally $k$, (or $k_{nc})$ in the present case, dictates the spatial geometry of the universe and NC effect can affect the spatial geometry in a non-trivial way.
One important aspect of these NC corrections needs to be taken care of since they inherently bring in anisotropy and inhomogeneity. To incorporate these fluctuations in generalized FRW models consistently some spatial  averaging hypothesis or prescription (over Lagrangian coordinates) is required. For example in a series of works Buchert \cite{av} has developed an averaging prescription as back reaction effects, that might be adopted for our case. Indeed, it is too early in our analysis to comment on the possibility of observing noncommutative effects in the context of cosmology from a numerical feasibility point of view.

\section{Summary and future prospects} Let us  summarize our work. In the first part we have provided a new framework to consider a generalization of a perfect fluid model in noncommutative space. The NC fluid model is developed from first principles. We have introduced the noncommutative algebra in particle coordinates (or equivalently Lagrangian degres of freedom) for the fluid. Subsequently we have derived the noncommutative algebra for fluid degrees of freedom in Eulerian picture. We have ensured that the new algebra satisfies Jacobi identities.   We have shown that the NC fluid model constructed here satisfies total mass (or number $N$) and total energy conservation principles. In our scheme expressions for number density and energy density remain canonical but noncommutative corrections are manifested in the number current and energy fluxes. Total momentum is conserved in the order of approximation we are interested in. Total angular momentum is also conserved in our level of accuracy although locally the rotational invariance is not preserved due to the introduction of the constant vectorial NC parameter $\theta^{ij}$. 

In the second part we have discussed how the present noncomutative fluid model can influence  conventional cosmological principles since the all important continuity and Euler equations get modified in a non-trivial way. In particular noncommutative corrections introduce inhomogeneity and anisotropy through cosmological perturbations.

Various directions of study open up from the present analysis. Let us list some of them.\\
{\it{Noncommutative fluid theory}}: A natural extension of the present analysis is to consider fluids with non-vanishing vorticity.\\
Our framework is essentially Hamiltonian but in conventional fluid dynamics there exists an action principle constructed in terms of Clebsch variables. It will be worthwhile to find a Clebsch parameterization for the noncommutative fluid.\\
The present study is restricted to a non-relativistic setup. Study of Schwinger's condition, a hallmark of relativistic quantum field theory, for non-relativistic fluid models has received attention recently in \cite{arpan}. Similar aspects can be analyzed for the present fluid system as well. Another interesting theme is the entropy consideration. Infact it is possible that already the noncommutative effects has brought in an entropy-variable since as we have pointed out it imposes some form of non-adiabeticity in the formalism.  It will also be very interesting if the noncommutative fluid can be generalized to a relativistic scenario. Works are in progress along these directions.\\
{\it{Noncommutative fluid in cosmological context}}: In the present work we just indicated how noncommutative space can have influence cosmological context without providing any explicit computations and results. The specific effects of noncommutativity-induced density perturbations, derived here, can be studied in detail.\\
{\it{Noncommutative fluid and gauge/gravity duality}}: Exploiting the  AdS/CFT duality the pioneering works \cite{gg} in the context of gauge/gravity duality  discuss the correspondence between relativistic,
conformal hydrodynamics and Einstein's theory of gravity. The fundamental role in this context is played by the energy-momentum tensor. The ideal fluid stress tensor does not contain derivatives of velocity and so can not incorporate dissipative effects. We have pointed out that the noncommutative effects can generate derivative terms in stress tensor components.  Thus extension of fluid dynamics to noncommutatve  fluid dynamics can lead to new features in gauge/gravity correspondence.
\vskip .5cm
\textbf{Acknowledgement}: The work of  P. D. is supported by INSPIRE, DST, India.

 \newpage

	There are several corrections related to our paper, \\
	Title:	Noncommutative Geometry and Fluid Dynamics; Authors:	Praloy Das, Subir Ghosh. \\ DOI:	10.1140/epjc/s10052-016-4488-8.\\
	The correct form of (17) is,
	\begin{eqnarray}
		\{j^i({\bf r}),j^j({\bf r^\prime})\}=j^j({\bf r})\partial_i\delta({\bf r}-{\bf r^\prime})+ j^i({\bf r^\prime})\partial_j\delta({\bf r}-{\bf r^\prime}).
		\label{t1}
	\end{eqnarray}	
	There is a crucial change of sign in the Dirac bracket of (37),
	\begin{eqnarray}
		\{\rho({\bf r}),\rho({\bf r^\prime})\}=-\theta_{ij} \partial_i \rho({\bf r})\partial_j\delta({\bf r}-{\bf r^\prime})
		\label{t2}
	\end{eqnarray}
	with (37,38) remaining unchanged. This leads to a canonical form of the  continuity equation,
	\begin{eqnarray}
		\dot{\rho}=\{H,\rho\}=-\partial_i(\rho v_i)
		\label{t3}
	\end{eqnarray}
	instead of (54). However the noncommutative extension of the  Euler equation (56) remains unchanged. Subsequently, contrary to our results of a noncommutativity-modified Friedmann equation in (84), in the present model, to first order in noncommutative parameter $\theta_{ij}$, the Friedmann equation remains unmodified.
	
	We point out some future directions where work is in progress:\\
	(i) Cosmological perturbations can manifest noncommutative corrections via our modified Euler equation.\\
	(ii) Higher order  $\theta_{ij}$ terms can modify the Friedmann equation.\\
	(iii) Nocommutative extension of fluid action, with a somewhat distinct bracket structure than the one presented here, can yield corrections in both continuity and Euler equations.
	
	We thank Rabin Banerjee and Arpan Krishna Mitra for pointing out the errors.

\end{document}